\documentclass[reprint,amsmath,amssymb,aps]{revtex4-1}

\usepackage{graphicx}% Include figure files
\usepackage{bm}% bold math

\begin{document}
%\widetext

\title{Scaling of the physical properties in Ba(Fe,Ni)$_2$As$_2$ single crystals : evidence for quantum fluctuations}

\author{P. Rodi\`ere$^1$,  T. Klein$^1$, L.Lemberger$^1$, K.Hasselbach$^1$, A.Demuer$^2$, J.Ka\v{c}mar\v{c}ik$^3$, Z.S.Wang$^{1,4}$, H.Q.Luo$^4$, X.Y.Lu$^4$ and H.H.Wen$^{4,5}$,  F. Gucmann$^6$, and C. Marcenat$^7$}
\address{$^{1}$ Institut N\'eel, CNRS and Universit\'e Joseph Fourier  BP166, F-38042 
Grenoble, France}
\address{$^2$ CNRS-LNCMI, UPR 3228, UJF-UPS-INSA, 38042 Grenoble, France}
\address{$^3$ Centre of  Low Temperature  Physics IEP  SAS \& FS-UPJ\v S, Watsonova 47, 043 53 Ko\v{s}ice, Slovakia}
\address{$^4$ Beijing National Laboratory for Condensed Matter Physics, Institute of Physics, Chinese Academy of Science,  Beijing 100190, China }
\address{$^5$ Center for superconducting Physics and Materials, National Laboratory of Solid State Microstructures and Department of Physics, Nanjing University, Nanjing  210093, China}
\address{$^6$ Institute of Electrical Engineering, Slovak Academy of Sciences, Dubravska cesta 9, 841 04 Bratislava, Slovakia}
\address{$^7$ SPSMS, UMR-E9001, CEA-INAC/ UJF-Grenoble 1, 17 rue des 
martyrs, 38054 Grenoble, France} 

\date{\today}

\begin{abstract}
We report on local magnetization, tunnel diode oscillator, and specific heat measurements in a series of Ba(Ni$_x$Fe$_{1-x}$)$_2$As$_2$ single crystals ($0.26\leq x\leq0.74$). We show that the London penetration depth $\lambda(T)=\lambda(0)+\Delta\lambda(T)$ scales as $\lambda(0)\propto 1/T_c^{0.85\pm0.2}$, $\Delta\lambda(T) \propto T^{2.3 \pm 0.3}$ (for $T<T_c/3$) and $\partial\Delta\lambda/\partial T^2\propto 1/T_c^{2.8\pm0.3}$ in both underdoped and overdoped samples. Moreover, the slope of the upper critical field ($H'_{c2}=-(dH_{c2}/dT)_{|T\rightarrow T_c}$)  decreases with $T_c$ in overdoped samples but increases with decreasing $T_c$ in underdoped samples. The remarkable variation of $\lambda(0)$ with $T_c$ and the non exponential temperature dependence of $\Delta\lambda$  clearly indicates that pair breaking effects are important in this system. We  show  that the observed scalings strongly suggest that those pair breaking effects could be associated with quantum fluctuations near 3D superconducting critical points.
\end{abstract}

\maketitle

\section{Introduction}

The discovery of superconductivity up to 55K in iron-based pnictides \cite{Kamihara} has generated
tremendous interest. Even though the symmetry of the order parameter in this multi-band system is not yet determined with certainty, a popular model is based on 
magnetic fluctuations associated with a  sign reversal of the order parameter between the hole and electron sheets of the Fermi surface (so called $s\pm$ model \cite{Mazin}). 
It has then been suggested by V.G.Kogan \cite{KoganI,KoganII} that the critical temperature ($T_c$)  could be strongly suppressed not  only by scattering breaking the time reversal symmetry (spin-flip scattering) but by any scattering mechanism. As a consequence,  for $T_c  << T_{c,0}$ ($T_{c,0}$ being the critical temperature in the absence of scattering) and an average of the order parameter over the Fermi surface being equal to zero, the superfluid density ($\rho_s$) is expected to vary as:
\begin{equation}
\rho_s \propto 1/\lambda^2 \propto T_c^2-T^2
\end{equation}
leading to a London penetration depth ($\lambda$) scaling as : $\lambda(0)  \propto 1/T_c$ and $\Delta\lambda(T) = \lambda(T)-\lambda(0) \propto T^2/T_c^3$ (for $T \rightarrow 0$). Moreover, the slope of the upper critical field close to $T_c$, $H'_{c2}=-(dH_{c2}/dT)_{|T\rightarrow T_c}$ is then expected to be proportional to  $T_c$  and finally the specific heat jump at $T_c$,  $\Delta C_p \propto T_c^3$. Some indications for the  $\Delta\lambda(T)$ \cite{GordonI} or $H'_{c2}$ and $\Delta C_p$ \cite{Budko} scalings have been observed in various pnictides  but a systematic analysis of the evolution of all those quantities on a given system was still lacking. 

Moreover,  the dependence of $\rho_s(0)$ on $T_c$ remains controversial. It has been initially suggested \cite{Carlo} that $\rho_s(0)$ could scale as $T_c$ but, measurements in overdoped Ba(Co$_x$Fe$_{1-x}$)$_2$As$_2$ rather suggested that $\rho_s(0)$ could be either proportional to $T_c^{2}$ \cite{Williams} or, on the contrary, almost $T_c$ independent  \cite{GordonII}. Those later measurements also indicated a strong reduction of the superfluid density  in underdoped samples, which has  been attributed to the coexistence of superconductivity and magnetism. 

It is worth noting that somehow contradictory results have also been observed in cuprates, emphasizing the fact that the $\rho_s(0)$ vs $T_c$ dependence can be very sensitive to the sample quality and/or dimensionality \cite{Hetel}. As in pnictides, initial measurements suggested that $\rho_s(0) \propto T_c$ in samples close to optimal doping \cite{Uemura} and this dependence found a straightforward explanation in this quasi 2D system assuming that $T_c$ is close to the Kosterlitz-Thouless-Berezinskii transition temperature $T_c \sim T_{KTD} = \Phi_0^2t/8\pi\mu_0\lambda^2$ ($t$ being the interlayer spacing). However, strong deviations from this behavior have been observed in highly underdoped samples for which $\rho_s(0)$ was found to scale as $T_c^\zeta$ with $1.6<\zeta<2.3$ in thick films \cite{Zuev} and ultraclean crystals \cite{Broun,Liang} or $\zeta \sim1$ in very thin films \cite{Hetel}. This change in $\zeta$ from $\sim 2$ to $1$ has been attributed to a dimensional crossover associated to the proximity of a quantum critical point \cite{QCP}.

We present here a detailed analysis of the doping dependence of $T_c$, $\lambda(0)$, $\Delta\lambda(T)$ and specific heat  in the Ba(Ni$_x$Fe$_{1-x}$)$_2$As$_2$ system. The measurements have been carried out in a series of Ba(Ni$_x$Fe$_{1-x}$)$_2$As$_2$ single crystals grown by self flux method. Details on the sample elaboration are given in \cite{Chen}. We will show that $\lambda(0)\propto 1/T_c^{0.85\pm0.2}$, $\Delta\lambda(T) \propto T^{2.3 \pm 0.3}$ (for $T<T_c/3$) and $\partial\Delta\lambda/\partial T^2_{|T\rightarrow 0}\propto 1/T_c^{2.8\pm0.3}$ in both underdoped and overdoped samples. Moreover, $H'_{c2}$ decreases with $T_c$ in overdoped samples but increases with decreasing $T_c$ for underdoped samples. Those results strongly suggest the presence of pair breaking effects and we will show that they can be consistently described assuming that those pair breaking effects are associated with the proximity of superconducting quantum critical points.

\section{London penetration depth}
\subsection{Lower critical field measurements}

The local  field has been measured by placing the samples on miniature GaAs-based quantum well Hall sensors. The external field was increased up to $H_a$ and  swept back to zero in order to measure the remanent field  ($B_{rem}$) trapped in the sample after the field cycle. All measurements were performed for $H_a\|c$. In the Meissner state, the external field is fully screened out and $B_{rem}$ is equal to zero. Vortices start to penetrate into the sample for $H_a > H_p$ (the first penetration field) and remain partially pinned in the sample when the field is swept back to zero, leading to a finite $B_{rem}$ value (see Fig.1a). For $H_a>H_p$, the remanent fields increases approximatively as $(H_a-H_p)^{1/\gamma}$ with $1/\gamma \sim 1.5-2.5$ and $H_p$ as hence been obtained by linearly extrapolating $B_{rem}^\gamma$ to zero (see Fig.1a). In order to avoid spurious effects associated to strong pinning preventing the vortex diffusion to the center of the sample (and hence the absence of any signal on probes located close to the center of the sample) $H_p$ has been measured on several locations with an array of miniature probes. As expected the measured $H_p$ value (slightly) increases as the distance between the probe and the sample edge increases. The $H_p$ values reported in Table 1 have been obtained with probes located at $\sim 5$ to $10 \mu$m from the edge and we estimate to $\sim 20$ \% the possible overestimation of this field.  

 \begin{figure}
\begin{center}
\resizebox{0.41\textwidth}{!}{\includegraphics{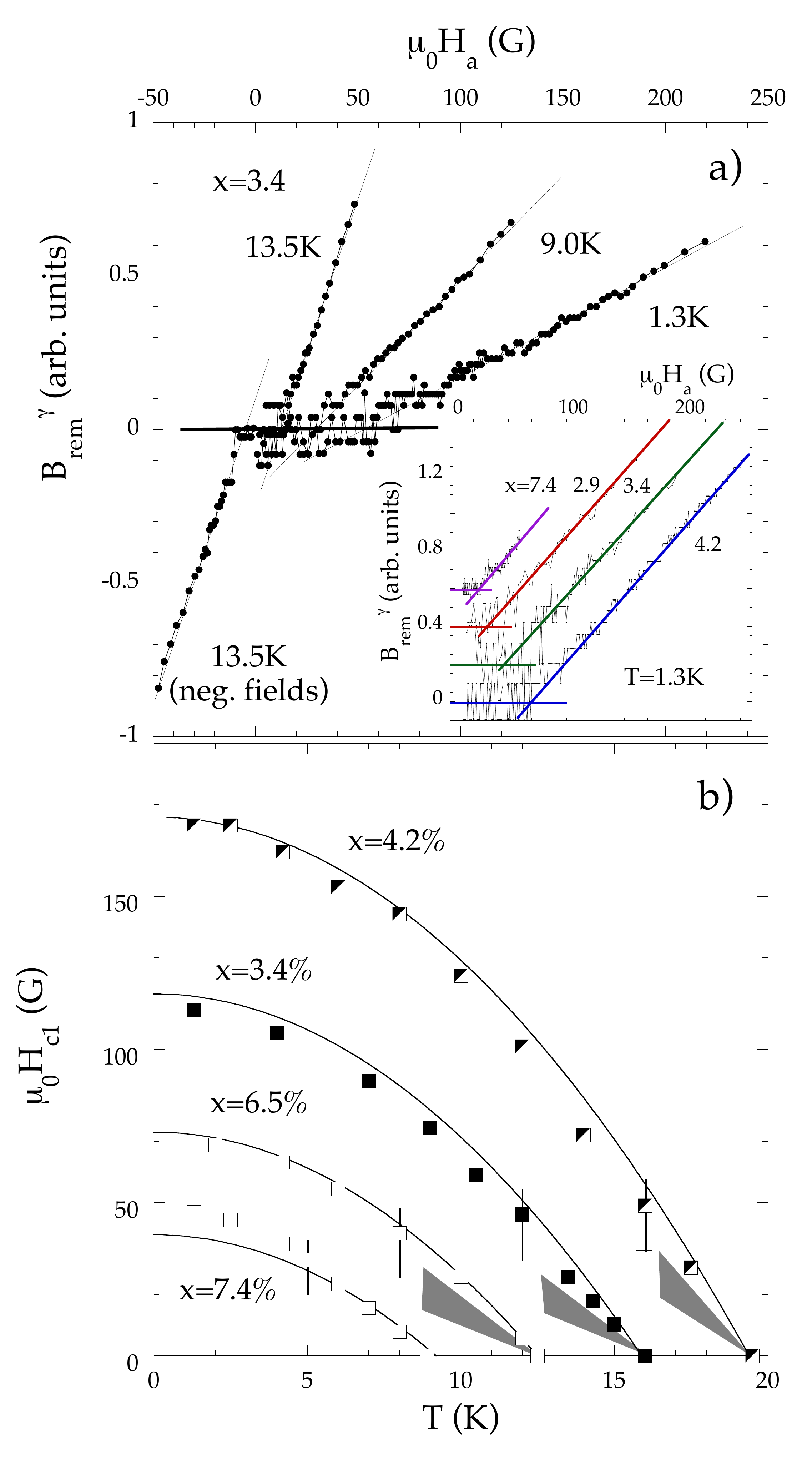}}
\caption{(a) $B_{rem}^\gamma$ as a function of the applied field for the indicated temperature in Ba(Ni$_x$Fe$_{1-x}$)$_2$As$_2$  with $x=3.4$\% ($B_{rem}$ being the remanent field trapped in the sample after a field excursion up to $H_a$ - see text for details - and $\gamma$ is an exponent $\sim 0.4-0.6$). The first penetration field is deduced from the linear extrapolation of $B_{rem}^\gamma$ to zero. Inset (color online) : Same as main panel for different $x$ values at $T=1.3$K (the different curves are shifted vertically by $0.2$ for clarity). (b) Temperature dependence of the lower critical field $H_{c1}$ in Ba(Ni$_x$Fe$_{1-x}$)$_2$As$_2$ single crystals (for the indicated doping  concentrations $x$). The solid lines are the values expected from Eq.(3) introducing only one free parameter (corresponding to the absolute value at zero $T$) for the whole set of data (see text for details). The shaded cones correspond to the values expected from the amplitude of the specific heat jump at $T_c$ (see Fig.4) and the slope of the upper critical field (see Fig.5 and text for details).}
\label{Fig.1}
\end{center}
\end{figure}

\begin{figure}
\begin{center}
\resizebox{0.47\textwidth}{!}{\includegraphics{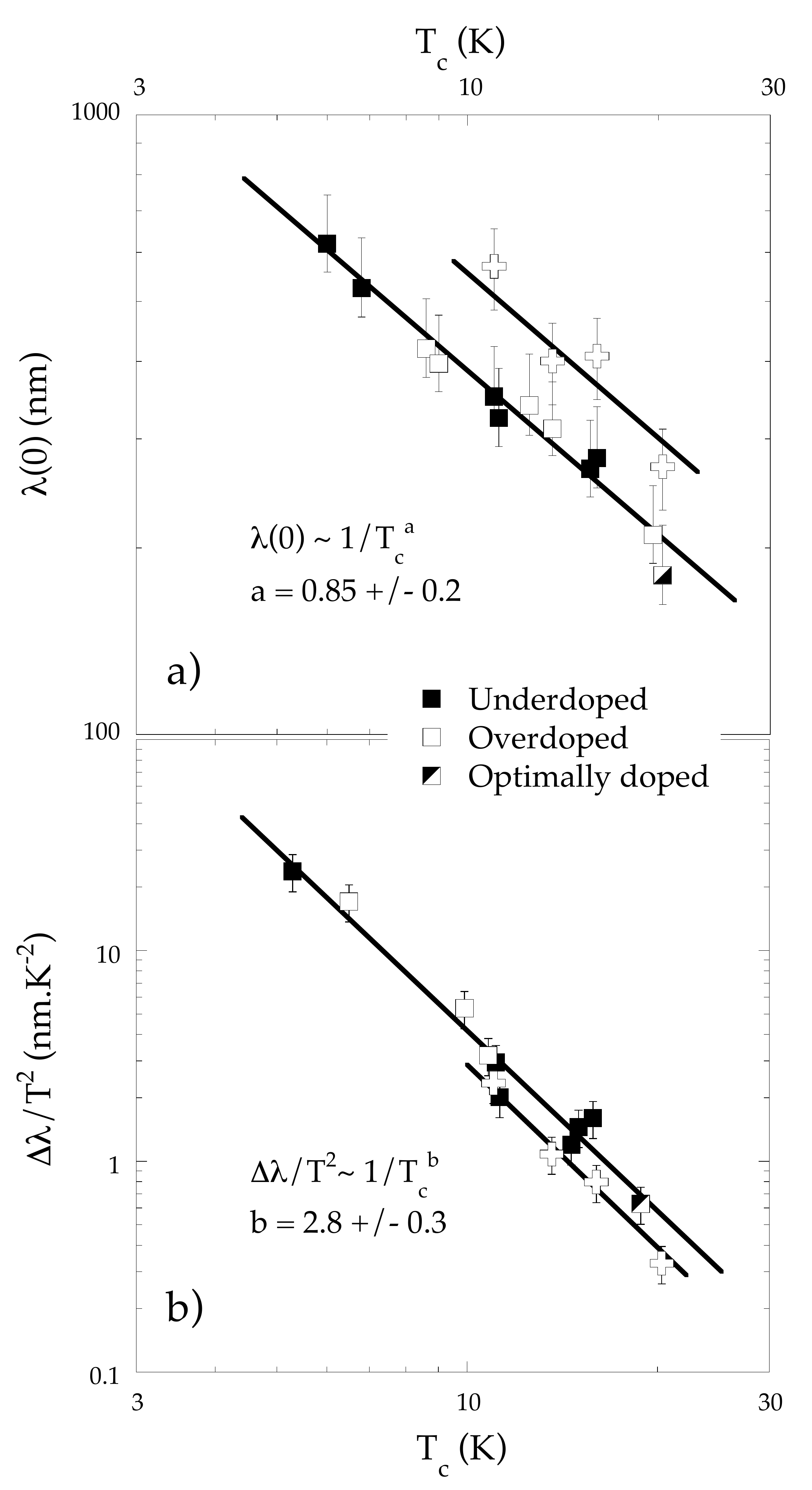}}
\caption{(a) London penetration depth at $T\rightarrow 0$ ($\lambda(0)$) as a function of the critical temperature $T_c$ in Ba(Ni$_x$Fe$_{1-x}$)$_2$As$_2$ crystals deduced from $H_{c1}$ measurements (squares, see Fig.1) and specific heat measurements (crosses, see text for details). The thick line is a guide to the eyes varying as $1/T_c^0.85$. (b) Slope of the $\Delta\lambda$ vs $T^2$ curve (see Fig.3) as a function of $T_c$ in samples from the same batch deduced from TDO measurements (squares). The crosses correspond to $\lambda(0)/2T_c^2$ (see text for details) introducing the $\lambda(0)$ values deduced from $C_p$ measurements (see upper panel).}
\label{Fig.2}
\end{center}
\end{figure}

This first penetration field is smaller than the lower critical field ($H_{c1}$) due to the expulsion of the flux lines for $H_a<H_p$ which increases the local field in the vicinity of the sample edges so that $H_{c1}=\alpha H_p$ with $\alpha > 1$. In the presence of geometrical barriers  (GB) \cite{Brandt}, $\alpha_{GB} \approx tanh(\beta d /w)^{-1/2} $ where $\beta$ varies from 0.36 in strips to 0.67 in disks ($d$ and $w$ being the thickness and width of the sample, respectively) whereas for elliptical samples (i.e. without geometrical barriers) the standard "demagnetization" factor $\alpha_{elliptical}=1/(1-N) \propto w/d$. All samples were chosen to present very similar aspect ratios (see Table 1) leading to very similar  corrections for the whole series whatever the origin of $\alpha$. In the following, we approximated the samples by disks leading to $2.4<\alpha_{GB} < 2.8$ from one sample to the other (see Table 1). However, it is important to note that a standard "demagnetization" correction  would lead to $2.7<1/(1-N)<3.3$ for our $d/w$ values ($\sim 1/4$) and would hence only lead to an overestimation of $H_{c1}$ by $\sim 12 \%$ i.e. an underestimation of $\lambda$ of $\sim 6 \%$. In the following, to evaluate the error bars, the uncertainty on $\alpha$ is taken  to $\pm 20$\%. The temperature dependences of  $H_{c1}$  (for the indicated $x$ values) are displayed in Fig.1b. 

 \begin{figure}
\begin{center}
\resizebox{0.49\textwidth}{!}{\includegraphics{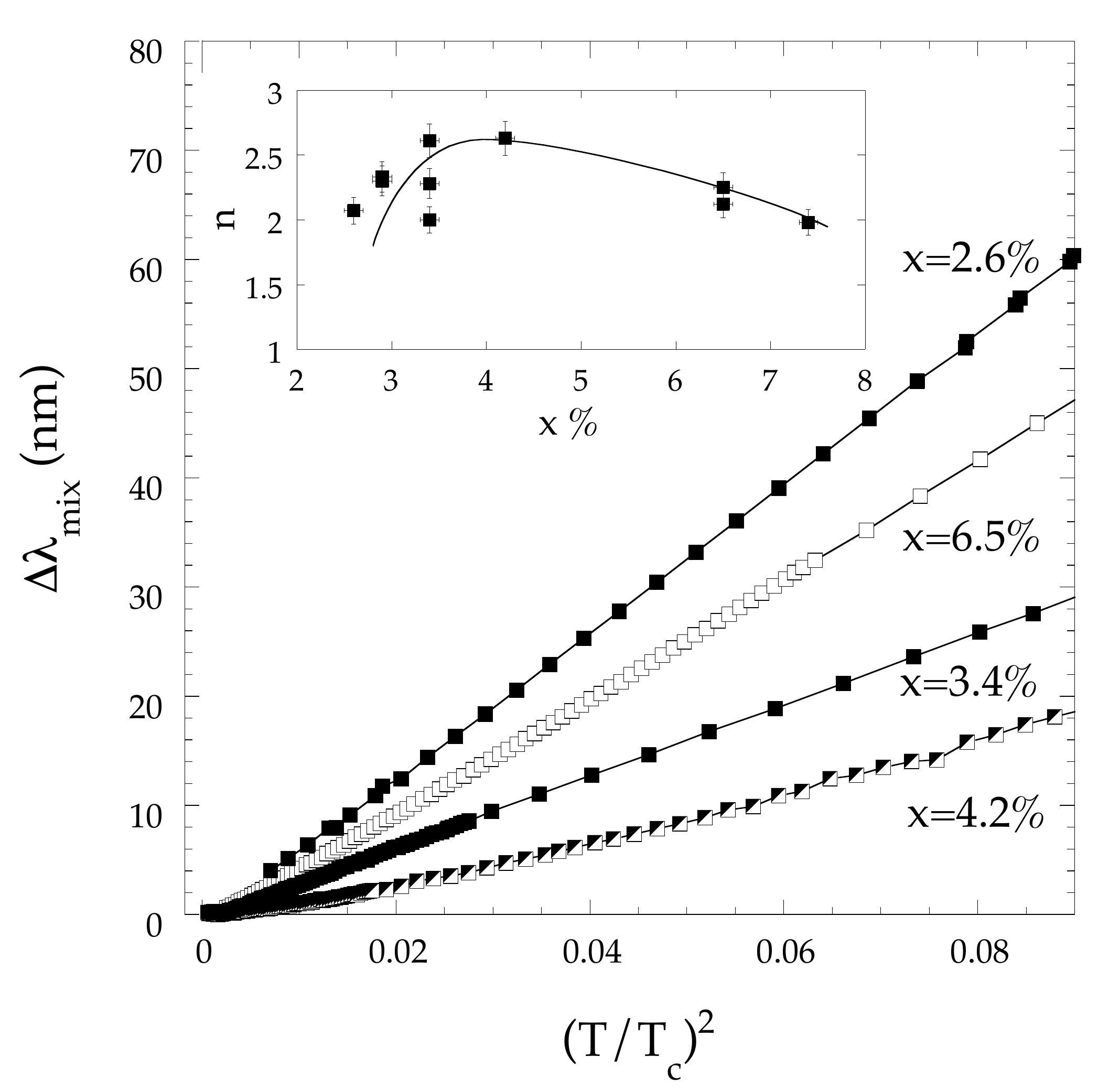}}
\caption{London penetration depth $ \lambda_{mix}=\lambda_{ab}+(d/w)\lambda_c$ ($ \sim \lambda_{ab}$ for $w>>d$) as a function of $(T/T_c)^2$ in Ba(Ni$_x$Fe$_{1-x}$)$_2$As$_2$ single crystals for the indicated $x$ values. As in Fig.2, closed, half-closed and open symbols refer to underdoped, optimally doped and overdoped samples, respectively. Inset : best fit values for the exponent n, assuming that $\Delta\lambda \propto T^n$.}
\label{Fig.3}
\end{center}
\end{figure}

 \begin{figure}
\begin{center}
\resizebox{0.48\textwidth}{!}{\includegraphics{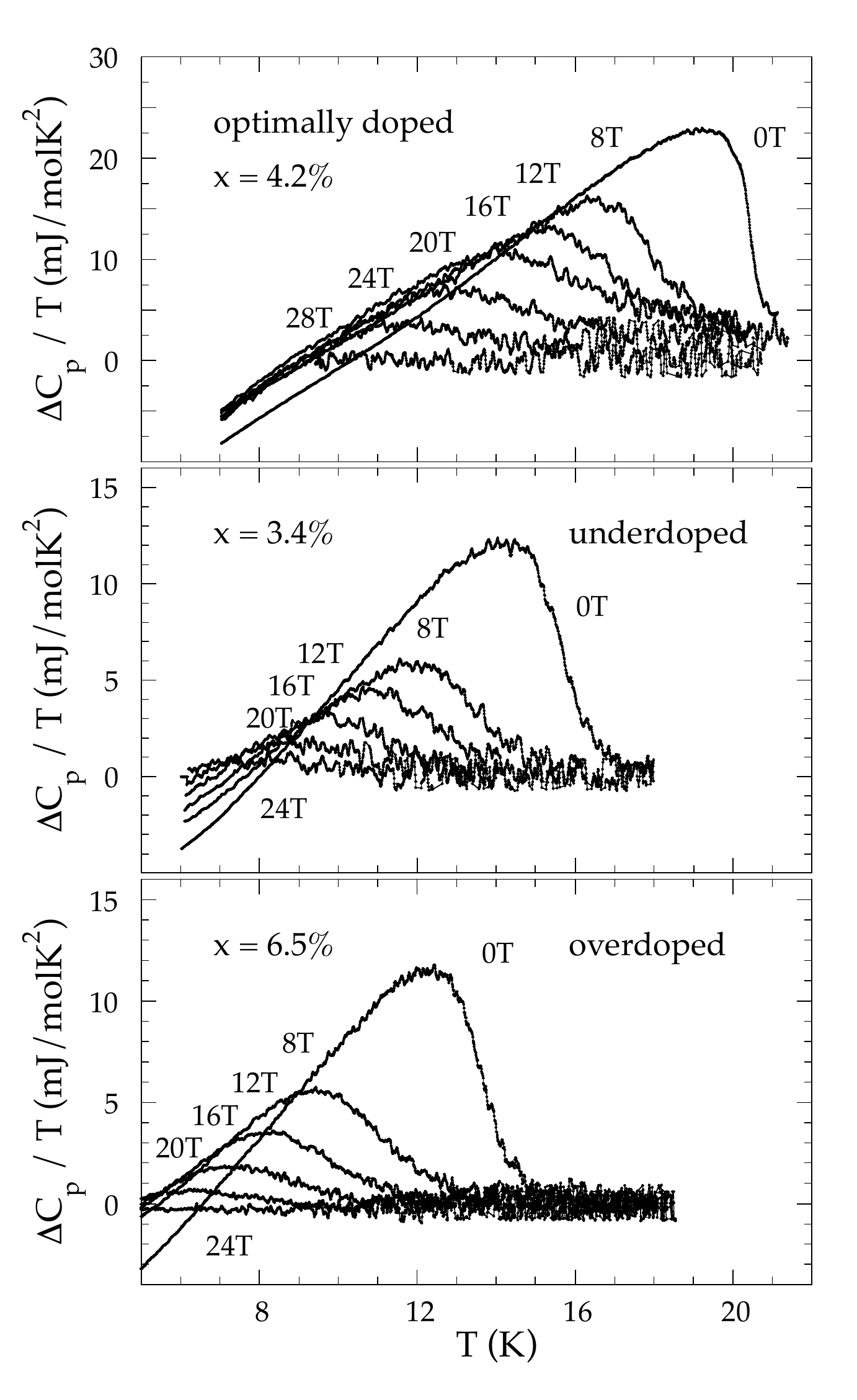}}
\caption{Temperature dependence of the field dependent part of the specific heat for the indicated field values in Ba(Ni$_x$Fe$_{1-x}$)$_2$As$_2$.}
\label{Fig.4}
\end{center}
\end{figure}

\begin{table*}
\caption{\label{table2} $x$, $d$, $w$ and $l$ are the composition, thickness, width and length of the Ba(Ni$_x$Fe$_{1-x}$)$_2$As$_2$ samples, respectively. The first penetration field $H_p$ has been defined as the field above which a remanent field is observed in the sample and the lower critical field $H_{c1}$ (see Fig.1) is deduced from $H_p$ introducing demagnetization corrections due to geometrical barriers ($\mu_0H_{c1}[G] =\alpha_{GB}\times \mu_0H_p[G]$ (see text). $\lambda$ is the London penetration depth (the values between brackets are deduced from specific heat measurements) and $\beta=\partial\lambda/\partial T^2$ (see Fig. 4). $H_{c2}$ is the upper critical field (see Fig.5, the values deduced from the irreversibility field are marked by an $^*$). All values are given for $T \rightarrow 0$. $\Delta C_p/T_c$ is the amplitude of the specific heat jump at $T=T_c$ (in mJ.mol$^{-1}$.K$^{-2})$. $G_i$ is the Ginzburg number (see text) and $T_c$ the critical temperature. Measurements techniques : HP=Hall Probe, Cp=Specific Heat and TDO=Tunnel Diode Oscillator.}
\begin{ruledtabular}
\begin{tabular}{cccccccccccccc}
$x$&$d$($\mu$m)&$w$($\mu$m) &$l$($\mu$m)&$\mu_0H_{c1}\sim\alpha_{GB}\times \mu_0H_p$&$\lambda$(nm)&$\beta$(nm/K$^2$)&$\mu_0H_{c2}$(T)&$\kappa$&$\Delta C_p/T_c$&Gi(10$^{-4}$)&$T_c$(K)&measured by\\
\hline
2.6 & 70 & 260 & 390 & 30 $\sim$ 2.8 $\times$ 11 & 530 & - & 23$^*$ & 150 & - & 4.0 & 6.8 & HP \\
2.6 & 140 & 340 & 450 & 25 $\sim$ 2.6 $\times$ 09 & 620 & - & - & - & - & - & 6.0 & HP \\
2.6 & 20 & 1040 & 1420 & - & - & 24 & - & - & - & - & 5.3 & TDO \\
\hline
2.9 & 50 & 160 & 250 & 70 $\sim$ 2.6 $\times$ 27 & 350 (570) & - & 35 & 120 & 10-12 & 2.9 & 11.0 & HP+Cp \\
2.9 & 20 & 70 & 80 & 80 $\sim$ 2.6 $\times$ 30 & 330 & - & - & - & - & - & 11.2 & HP \\
2.9 & 5 & 450 & 570 & - & - & 2.9 & - & - & - & - & 11.1 & TDO \\
2.9 & 10 & 700 & 950 & - & - & 2.0 & - & - & - & - & 11.2 & TDO \\
\hline
3.4 & 90 & 280 & 320 & 100 $\sim$ 2.6 $\times$ 38 & 290 (410) & - & 46 & 100 & 14-16 & 2.9 & 16.0 & HP+Cp \\
3.4 & 60 & 210 & 290 & 110 $\sim$ 2.7 $\times$ 43 & 270 & - & - & - & - & - & 15.6 & HP \\
3.4 & 50 & 580 & 630 & - & - & 1.2 & - & - & - & - & 14.6 & TDO \\
3.4 & 5 & 330 & 640 & - & - & 1.4 & - & - & - & - & 15.0 & TDO \\
3.4 & 10 & 470 & 820 & - & - & 1.6 & - & - & - & - & 15.8 & TDO \\
\hline
4.2 & 70 & 300 & 340 & 240 $\sim$ 2.8 $\times$ 85 & 180 (270) & - & 52 & 80 & 24-28 & 1.7 & 20.2 & HP+Cp \\
4.2 & 10 & 45 & 45 & 180 $\sim$ 2.8 $\times$ 62 & 210 & - & - & - & - & - & 19.6 & HP \\
4.2 & 10 & 300 & 590 & - & - & 0.6 & - & - & - & - & 18.8 & TDO \\
\hline
6.5 & 90 & 280 & 380 & 80 $\sim$ 2.4 $\times$ 35 & 310 (400) & - & 28 & 90 & 13-15 & 1.2 & 13.5 & HP+Cp \\
6.5 & 100 & 250 & 300 & 70 $\sim$ 2.7 $\times$ 26 & 340 & - & - & - & - & - & 12.5 & HP \\
6.5 & 20 & 850 & 900 & - & - & 3.2 & - & - & - & - & 10.8 & TDO \\
6.5 & 20 & 310 & 630 & - & - & 5.3 & - & - & - & - & 9.9 & TDO \\
\hline
7.4 & 80 & 280 & 340 & 45 $\sim$ 2.6 $\times$ 17 & 420 & - & 13$^*$ & 80 & - & 1.3 & 8.6 & HP \\
7.4 & 50 & 190 & 220 & 50 $\sim$ 2.8 $\times$ 18 & 400 & - & - & - & - & - & 9.0 & HP \\
7.4 & 40 & 540 & 900 & - & - & 17 & - & - & - & - & 6.5 & TDO \\
\end{tabular}
\end{ruledtabular}
\end{table*}

Finally, $\lambda(0)$ has been deduced from the upper and lower critical fields writing : $\mu_0H_{c2}=\Phi_0/2\pi\xi^2$ (with $H_{c2}(0)\sim0.7.T_c.(dH_{c2}/dT)_{|T\rightarrow T_c}$ \cite{WHH}, see section III) and $\mu_0H_{c1}=(\Phi_0/4\pi\lambda^2).(Ln(\kappa)+c(\kappa))$  where $\kappa=\lambda/\xi$ and $c(\kappa)$ is a $\kappa$ dependent function tending towards $\sim 0.5$ for large $\kappa$ values.  $\kappa$ has hence been deduced from the $H_{c2}/H_{c1}$ ratio (see table 1) and the corresponding $\lambda (0)=\kappa\xi(0)$ values have been reported on Fig.2a as a function of $T_c$. As shown,  similar values ($\lambda(0) \sim 1/T_c^{0.85 \pm 0.2}$) are obtained on both sides on the superconducting dome in striking contrast with the result obtained by Gordon {\it et al.} in Ba(Ni$_x$Fe$_{1-x}$)$_2$As$_2$ \cite{GordonII}. Our measurements do hence {\it not} support the scenario of a sharp increase of $\lambda(0)$ due to the presence of a magnetic gap in underdoped samples but rather suggest that only a small fraction of the Fermi surface is affected by the antiferromagnetic coupling.

Note that $(dH_{c1}/dT)_{|T\rightarrow T_c}$ (and hence $\lambda(0)$) can also be deduced from specific heat measurements (see section III). Indeed, the amplitude of the specific heat  jump at $T_c$ is equal to : $\Delta C_p=(\mu_0T_c).(dH_c/dT)^2_{|T\rightarrow T_c}
\sim (\mu_0T_c).(dH_{c2}/dT)_{|T\rightarrow T_c}.(dH_{c1}/dT)_{|T\rightarrow T_c}/[ln(\kappa+0.5]. $
 Deducing $H_{c2}(T)$ from the shift of the anomaly under magnetic field (see section III.A), one directly obtains $(dH_{c1}/dT)_{|T\rightarrow T_c}$ and $\lambda^{C_p}(0)$ can then be calculated using the same procedure as above, assuming that $H_{c1}(0)\sim0.7.T_c.(dH_{c1}/dT)_{|T\rightarrow T_c}$. The corresponding $H_{c1}(T\rightarrow T_c)=(T-T_c)\times(dH_{c1}/dT)_{|T\rightarrow T_c}$ and $\lambda^{C_p}(0)$ values have been reported on Fig.1b (shaded cones) and Fig.2a (crosses), respectively. It is important to note that those values do not depend on any demagnetization correction. As shown in Fig.2a, the as-deduced $\lambda^{C_p}$ values are in reasonable agreement with  those deduced from $H_p$ ($\lambda^{H_p}$) measurements ($\lambda^{H_{p}}/\lambda^{C_p} \sim 0.7$) and both measurements lead to very similar dependences on $T_c$. 

The most remarkable feature is the very large variation of the superfluid density which varies over one order of magnitude for $T_c$ values ranging from $\sim 6.8$K to $\sim 20.2$K.  Such a strong dependence is unexpected in conventional superconductors, but strongly suggests the presence of  pair breaking effects. 
\subsection{Tunnel Diode Oscillator Technique}

The samples were glued at the end of a sapphire rod which was introduced in a coil of inductance $L$. The variation of the London magnetic penetration depth induces a change in $L$ and hence a shift of the resonant frequency $\delta f(T)=f(T)-f(T_{min})$ of a LC oscillating circuit (14MHz) driven by a Tunnel Diode. This shift, renormalized to the one corresponding to the extraction of the sample from the coil ($\Delta f_0$) is then equal to the fraction ($\delta V/V$) of the sample which is penetrated by the field. For $H\|c$, $\delta V$ is related to the in-plane penetration depth $\lambda_{ab}$ through some calibration constant depending on the sample geometry. However, this constant can be altered by edge roughness effects (see discussion in \cite{Klein}) and we have hence decided to perform all measurements with $H\|ab$. Indeed, the surfaces parallel to the $ab$-planes are much flatter and $\delta V/V$ is, in this case, directly given by $\delta V/V \sim 2(\lambda_{c}/w+\lambda_{ab}/d)=\delta V/V  \sim 2/d\times[\lambda_{ab}+(d/w)\lambda_c]$ without any geometrical correction ($\lambda_c$ being the penetration depth parallel to the c-axis). In contrast to $H_{c1}$ measurements for which we used rather "thick" ($d/w \sim 1/4$) samples in order to reduce the uncertainty related to geometrical corrections, we have, in this case, selected very thin samples ($d/w<<1$) (see Table 1) so that $\lambda_{ab}+(d/w)\lambda_c = \lambda_{mix} \sim \lambda_{ab}$ (for weakly anisotropic systems \cite{anisotropy}). 

The temperature dependence of the penetration depth is clearly non exponential in all measured samples but can be well described by a power law : $\Delta\lambda_{mix}(T) = \lambda_{mix}(T)-\lambda_{mix}(0) = A.T^{\sim 2.3 \pm 0.3}$ for $T \leq T_c/3$ (see Fig.3). A very similar behavior has been reported in a large number of pnictides in both 1111 and 122 systems \cite{GordonI} (as well as in Fe(Se,Te), see \cite{Klein} and references therein).  This dependence is another clear indication for the presence of pair breaking effects but it is also important to note that the slope of the $\Delta\lambda$ vs $T^2$ curve, $\partial\Delta\lambda/\partial T^2_{|T\rightarrow 0}$ is proportional to $1/T_c^{2.8}$ (see discussion in section IV.A).

\section{Upper critical field}
\subsection{Specific heat measurements}
Finally, $C_p$ measurements have been performed in magnetic fields up to 28 T using an high 
sensitivity AC technique  (typically $1$ part in $10^3$). Heat was supplied to the sample by a light emitting diode via an optical fiber and the corresponding temperature oscillations were recorded with a thermocouple. In order to obtain quantitative $C_p$ values, special care has been taken in the calibration procedure to measure a copper standard and the addenda in the exact same conditions. For $x=3.4$, $4.2$ and $6.5$ \% well defined specific anomalies were obtained in zero field (see Fig.4) and  this anomaly progressively shifted with magnetic field. $T_{c2}$ has hence been defined as the temperature corresponding to the mid point of the transition for a given value of the external field. The transition became broader and weaker for $x=2.9$ \% (not shown) and finally too weak (and/or broad) to allow any accurate determination of $H_{c2}$ from specific heat measurements for $x=2.6$ \% and $x=7.4$ \%. The $H_{c2}(T)$ values have been reported in Fig.5 and the corresponding $H'_{c2}$ values are displayed in Fig.6b together with the $T_c$ versus $x$ data. 

 \begin{figure}
\begin{center}
\resizebox{0.48\textwidth}{!}{\includegraphics{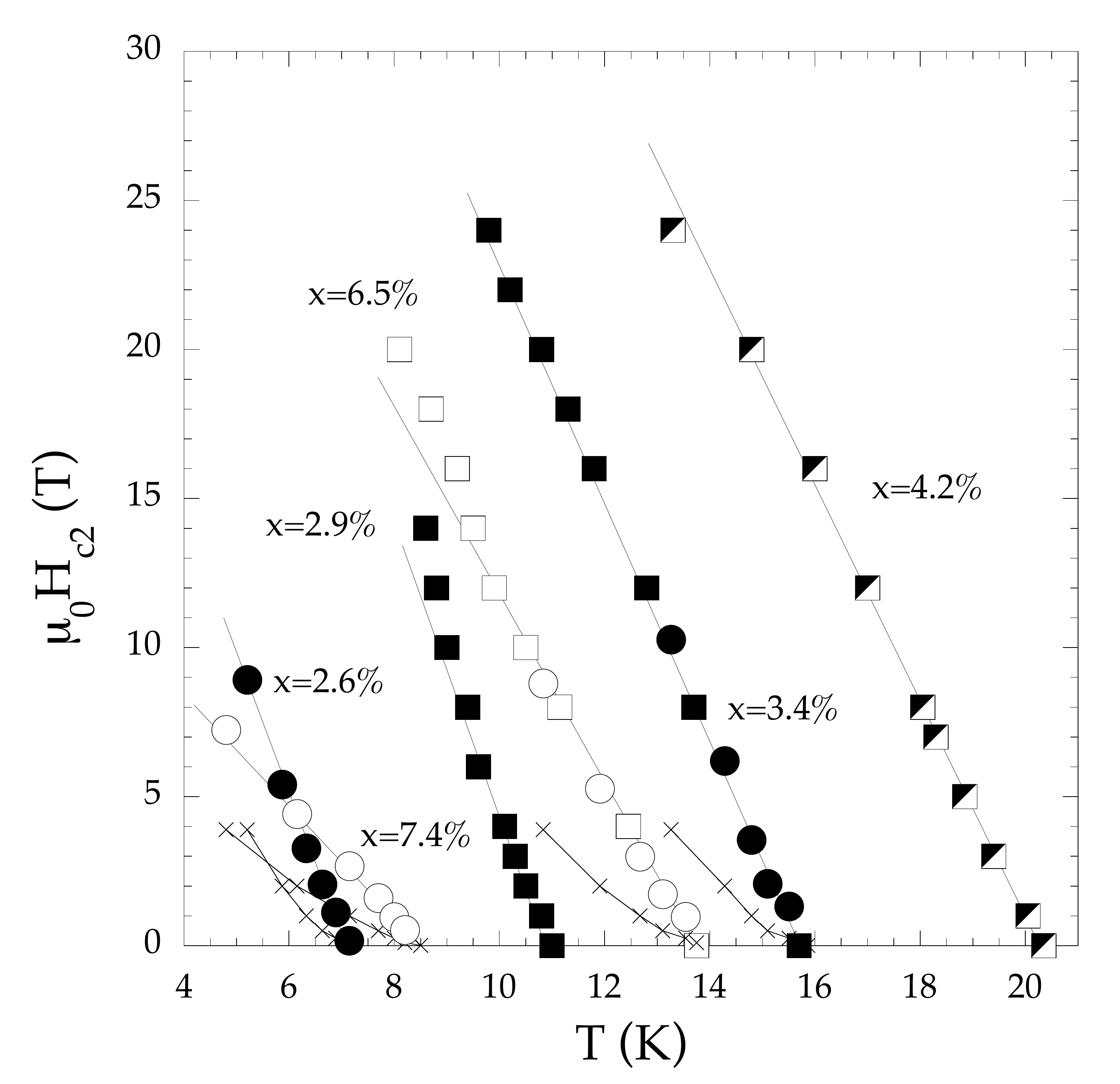}}
\caption{Temperature dependence of the upper critical field $H_{c2}$ deduced from the mid-point of the specific heat anomaly (squares, see also Fig.2 for symbol correspondance). The crosses correspond to the irreversibility field ($H_{irr}$) deduced from the onset of diamagnetic screening and circles to $H_{c2}$ values deduced from  $H_{irr}$ (see text for details).}
\label{Fig.5}
\end{center}
\end{figure}

\subsection{Transmittivity measurements}
In order to obtain the upper critical field for the two low $T_c$ samples, we performed transmittivity measurements ($T'_H$). The AC component of the local induction on the Hall probe was recorded in presence of an AC modulation field ($h_{ac}\sim 1G$, $\omega\sim 200$ Hz) and $T'_H$ has been defined as : $T'_H=\frac{(B_{AC}(T)-B_{AC}(T<<T_c))}{(B_{AC}(T>T_c)-B_{AC}(T<<T_c))}$. The irreversibility field ($H_{irr}$) was defined as the onset of diamagnetic response (i.e. locus of minimal observable screening current). Writing the Ginzburg-Landau free 
energy functional in terms of Lowest Landau Level (LLL)  eigenfunctions,  $H_{c2}$ can then be deduced from $H_{irr}$ through  \cite{Mikitik} : 
\begin{equation} 
(1-h)(1-t^{2})^{1/3}=(th)^{2/3}\times CGi^{1/3}
\end{equation}
where $t =T/T_{c}$, $h = H_{irr}(T)/H_{c2}(T)$, $C$ is a constant depending on the amount of disorder present in the sample and $Gi= \frac{1}{8}\left[ k_{B}T_{c}/\varepsilon_0\epsilon\xi(0)) \right]^{2}$ is the Ginzburg number (see Table 1) with 
$\varepsilon_0=(\Phi_0/4\pi\lambda(0))^2$ the vortex line energy and $\epsilon$ the anisotropy \cite{anisotropy}. 

 \begin{figure}
\begin{center}
\resizebox{0.48\textwidth}{!}{\includegraphics{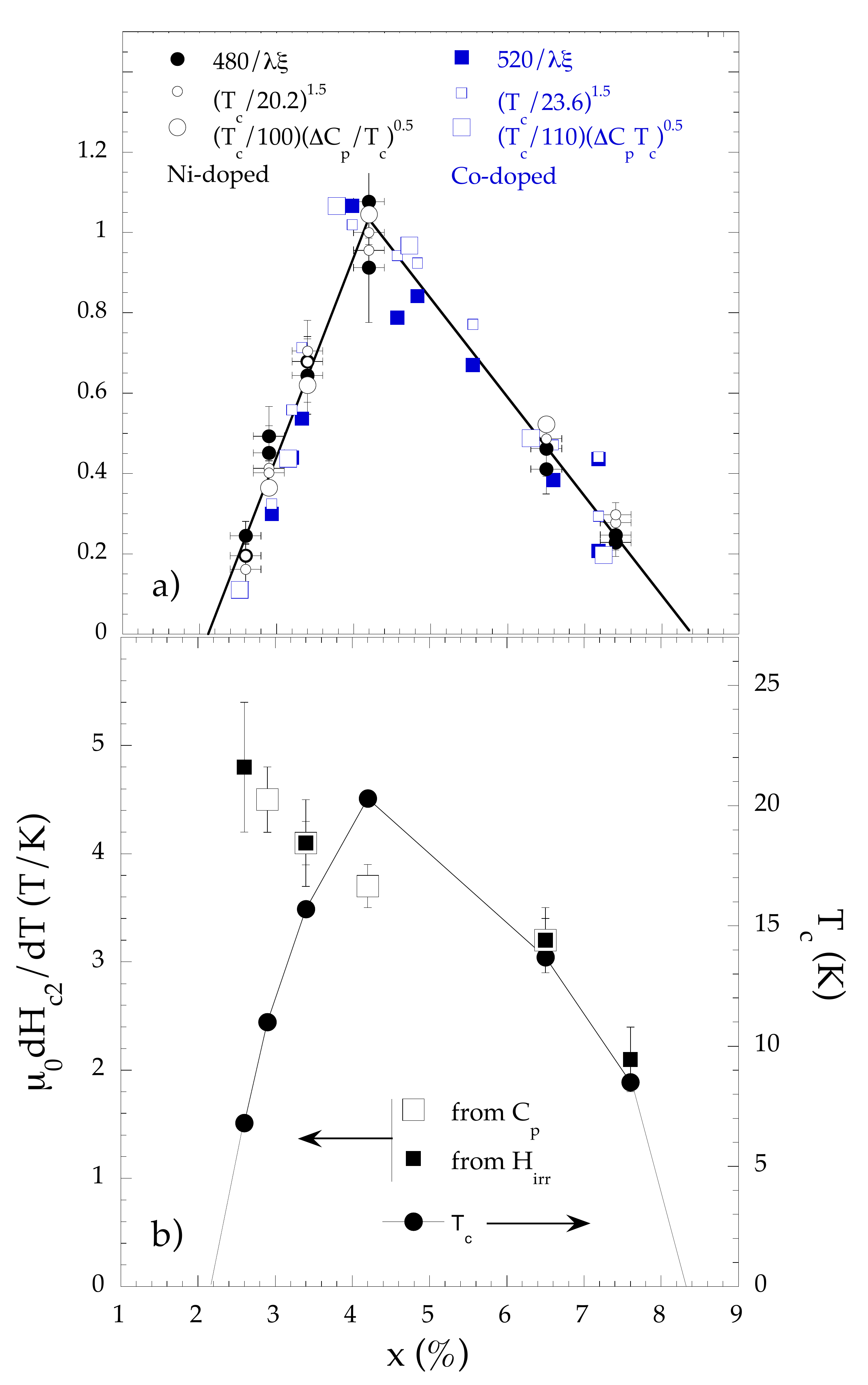}}
\caption{(color online) (a) $1/\xi\lambda$ (closed symbols), $T_c^{1.5}$ (small open symbols) and $T_c(\Delta C_p/T_c)^{0.5}$ (large open symbols) as a function of doping in both Ba(Fe$_x$,Ni$_{1-x}$)$_2$As$_2$ (present work, (black) circles) and Ba(Fe$_x$,Co$_{1-x}$)$_2$As$_2$(blue) squares). The $x$ values in Co-doped samples have here been rescaled by a factor $\sim 0.6$. (b) Left scale : slope of the critical field $H'_{c2}=dH_{c2}/dT_{|T\rightarrow T_c}$ (solid lines in Fig.5b) as a function of the composition ($x$) in Ba(Ni$_x$Fe$_{1-x}$)$_2$As$_2$ single crystals where $H_{c2}$ has been deduced from specific heat measurements (open squares) and from the irreversibility field (solid squares). Right scale : critical temperature $T_c$ (solid circles) as a function of $x$.}
\label{Fig.6}
\end{center}
\end{figure}

As shown in Fig.5, the $H_{c2}(T)$ values  (diamonds) deduced from $H_{irr}(T)$ (crosses) taking $CG_i^{1/3} \sim 1$  are in very good agreement with the values deduced from the specific heat measurements (squares) for both $x=3.4$ and $x=6.5$ \%. Note that the irreversibility line lies significantly below the $H_{c2}(T)$ line, clearly suggesting the presence of vortex liquid phase in this systems \cite{Pribulova}. The corresponding $C$ value ($\sim 15$) is very close to the one obtained in (Ba,K)Fe$_2$As$_2$ \cite{Kacmarcik} and we have calculated the $H_{c2}$ values for $x=2.6$ and $7.4$ \% using this $C$ value (i.e. taking $CG_i^{1/3} \sim 1.1 \pm 0.2$ and $CG_i^{1/3} \sim 0.8 \pm 0.2$ for $x=2.6$ and $x=7.4$, respectively) We hence confirmed the decrease of $H'_{c2}$ with $T_c$ in  overdoped samples and the increase of $H'_{c2}$ for $T_c \rightarrow 0$ for low doping contents (see also Fig.6b). Note that a similar behavior has been reported by Vinod {\it al.} \cite{Vinod} in Ba(Ni$_x$Fe$_{1-x}$)$_2$As$_2$. Even though the determination of $H_{c2}$ from transport measurements may be altered by fluctuation effects, those measurements also suggested that $H'_{c2}$ could be proportional to $T_c$ in overdoped samples and that the $H'_{c2}/T_c$ ratio increases in underdoped samples. Finally, note also that a different scaling has been recently obtained in overdoped BaFe$_2$(As,P)$_2$ samples in which $H'_{c2}$ rather scales as $T_c^2$ \cite{Chaparro} (see discussion in section IV.B). 

\section{discussion and concluding remarks}

In summary, we have shown that : 

i) the penetration depth $\lambda$ strongly varies with the critical temperature of the sample,

 ii) $\lambda(0)$ scales as $T_c^{0.85\pm0.2}$ for both underdoped and overdoped samples, 
 
 iii) the temperature dependence of $\lambda$ is non exponential but varies as $\Delta\lambda \propto T^{2.3\pm 0.3}$ for $T<T_c/3$ ,
 
 iv) the slope of the upper critical field decreases with decreasing $T_c$ in overdoped samples but increases with decreasing $T_c$ in underdoped samples. 
  
Points i) and iii) strongly suggest that pair breaking effects are important. As pointed out by V.G.Kogan \cite{KoganI,KoganII}, in the $s\pm$ model, the Cooper pairs are expected to be be very sensitive to all scattering events and the superconducting condensate is hence progressively destroyed on both side of the superconducting dome,  in agreement with the observation of a non residual Sommerfeld coefficient in specific heat data in Co-doped samples \cite{Hardy}. However, it is still unclear whether the system lies close to the critical disorder for which $T_c \rightarrow 0$ and whether the gap vanishes or not. Alternatively, those pair breaking effects can be associated with quantum fluctuations \cite{QCP}. Those fluctuations require a finite value of the coupling strength even for $T_c \rightarrow 0$ \cite{Ramazashvili}. They can hence not be observed in a standard BCS superconductors but can in the presence of pair breaking effects as superconductivity only develops above some finite value of the coupling strength in this case. The scaling properties associated with those two scenarios (strong pair breaking effects close to critical disorder and superconducting quantum fluctuations) are discussed below.

\subsection{Strong pair breaking effects for $<\Omega>=0$}

As discussed by V.G.Kogan \cite{KoganI,KoganII}, the superfluid density is expected to vary as  $(T_c(x)^2-T^2)$ (Eq.(1)) for an average of the order parameter over the Fermi surface $<\Omega>=0$ and vanishingly small critical temperatures (see also \cite{Vorontsov}). Writting $1/\lambda (T)^2=1/(\lambda(0)+\Delta\lambda(T))^2 \sim (1/\lambda(0)^2)\times(1-2\Delta\lambda (T)/\lambda(0))$ (for $\Delta\lambda<<\lambda(0)$) one then expects that $\Delta \lambda(T) \sim\beta T^2$ with $\beta =  \lambda(0)/2T_c^2$. The exponent $n$ slightly larger than 2 observed experimentally can be attributed to a non zero gap value   \cite{Vorontsov} but in order to compare the different samples within each other, we assumed in the discussion below that $n =2$. Neglecting the - small - variation of $ln(\kappa)$ with $x$ one also expects that  :
\begin{equation} 
H_{c1}(x,T)=K(x)\times(T_c(x)^2-T^2).
\end{equation}
This equation leads to a very satisfying agreement to the data introducing only {\it one single} adjustable parameter ($K(x)=K_0$) for the whole set of $x$ values (see solid lines in Fig.1b). As mentioned above, the aim is here not to obtain the best fit to the data (the agreement could for instance be improved by adjusting the exponent n) but to emphasize that the whole set of data can be reproduced in a very satisfactory way only assuming that Eq.(1) is valid. Eq.(1) also suggests that $\lambda(0) \propto 1/T_c$ in very reasonable agreement with the $\lambda(0)=A/T_c^{0.85}$ scaling observed in Fig.2a (point ii)). One then expects $\beta =\lambda(0)/2T_c^2 = B/T_c^{2.85}$ in very good agreement with the experimental data (see Fig.2b). TDO measurements hence also lead to $\lambda(0) \sim A/T_c^{0.85}$. The corresponding $A$ value is on the order of $ \sim 5600$ nmK$^{0.85}$ e.g. about two times larger than the one obtained from $H_{c1}$ measurements ($A \sim 2800$ nmK$^{0.85}$) but in very fair agreement with the one deduced from $C_p$ measurements  ($\sim 4000$ nmK$^{0.85}$, see corresponding $\lambda(0)/2T_c^2$ values in Fig.2b).  It is important to note that, this scaling law has been obtained through three  {\it independent} techniques (with an average $A$ value $\sim 4100$ nmK$^{0.85}$). Note that much larger $\beta$ values (from $\sim 1.2$ nm/K$^2$ in the optimally doped sample to $\sim 45$ nm/K$^2$ for $x=7.2$\%) have been obtained by Martin {\it et al.} \cite{Martin} in Ni doped crystals  but those values would correspond to $\lambda(0)$ ranging from $\sim 900$ nm to $\sim 1600$ nm (in the pair breaking scenario) i.e. much larger than any experimental value. 

The scaling properties of the magnetic penetration  depth are hence in very reasonable agreement with the strong pair breaking model suggested by V.G.Kogan \cite{KoganI,KoganII} and it is reasonable to attribute, in this scenario, the small deviations from the expected scaling laws to the non zero gap value. However, it is important to note that this model assumes a very strong reduction of $T_c$ suggesting that $T_{c,0}$ could largely exceed 100 K in pnictides in the absence of scattering which remains very puzzling. Moreover, sample dependent parameter (such as the density of states or the Fermi velocity) enter in the prefactors of the scaling functions and it is hence quite surprising to obtain one single $A$ value (i.e. $K$ value for the $H_{c1}$ data) for all doping contents \cite{ARPES}. Moreover, even though V.G.Kogan pointed out in \cite{KoganII} that the $H'_{c2} \propto T_c$ dependence expected for $<\Omega>=0$ would actually be reversed ($H'_{c2} \propto 1/T_c$) for  $<\Omega>\neq0$, the strong difference in $H'_{c2}(T_c)$ between underdoped and overdoped samples (point iv)) and {\it not} in $\lambda(0)$ vs $T_c$, can hardly be explained in the framework of this model.

\subsection{Proximity of a superconducting quantum critical point}

In the presence of quantum fluctuations, static and dynamical properties are inextricably mixed, so that the value of the dynamic exponent $z$ directly enters in the scaling properties of the thermodynamical properties \cite{QCP}. However, for d=2 ($d$ being the dimension), $z$ cancels out in the $\rho_s(0)$ vs $T_c$ scaling and $\rho_s(0) \propto T_c$  for all $z$. On the other hand, for d=3, $d+z \geq 4$ (i.e. larger than the Gaussian end point value) and the superfluid density is expected to vary as : $\rho_s(0) \propto |p-p_c|$
 where $p$ is the parameter driving the transition (here the doping content) and $p_c$ its critical value. The critical temperature varies as : $T_c \propto |p-p_c|^\Psi$ where the shift exponent $\Psi$ is related to $z$ through $\Psi =z/(z+1)$ \cite{QCP} and one expects :
 \begin{equation}
 \rho_s(0) \propto T_c^{(z+1)/z}
 \end{equation}
with $z=2$ in standard superconducting quantum critical points and $z=1$ in the presence of nodes in the superconducting gap \cite{Ramazashvili}. One then expects either $\rho_s(0) \propto T_c^{1.5}$ for $z=2$ or $\rho_s(0)\propto T_c^{2}$ for $z=1$. The change from $\rho_s(0)\propto T_c$ to $\rho_s(0)\propto T_c^2$ with the sample thickness in underdoped YBaCuO has hence been interpreted as a dimensional crossover from 2D thin samples to 3D samples with $z=1$ \cite{Hetel}. 
 
In our case, $\rho_s \propto T_c^{1.7\pm0.4}$ and it is hence impossible to distinguish between the $z=1$ and $z=2$ cases ($d=3$). On the other hand, $1/\xi_0^2$ is also expected to scale as $T_c^{1/\Psi}$ and, as  for the strong pair breaking model discussed above, it would be difficult to explain the different $H'_{c2}(T_c)$ dependences observed for underdoped and overdoped samples assuming that they are in the clean limit ($H'_{c2} \propto 1/\xi_0^2T_c$). This difference can however be understood in the dirty limit for which $\xi \sim (\xi_0l)^{0.5}$ where $l$ and $\xi_0$ are the mean free path and coherence length without disorder, respectively. A non symmetric $l(x)$ dependence would lead to different $H'_{c2}$ dependence for underdoped and overdoped samples. Note that some difference would also be expected for $\lambda \sim \lambda_0(\xi_0/l)^{0.5}$ ($\lambda_0$ being the penetration depth without disorder) but this difference is much weaker than for $H'_{c2}$ and remains within our error bars.

Interestingly, $l$ is expected to cancel out in the $\xi\lambda$ product which is equal to $\xi_0\lambda_0$ for both clean ($l>\xi_0$) and dirty ($l<\xi_0$) samples and one expects :
\begin{equation}
1/\xi\lambda\propto T_c^{(z+1)/z}\propto |x-x_c|
\end{equation}
 on both sides of the superconducting dome, independently of the sample quality. Note that, two quantum critical points are present in this scenario, corresponding to the two end points of the dome. As shown in Fig.6a, $1/\xi\lambda$  (closed circles) and $T_c^{3/2}$ (small open circles) present very similar dependence on the doping content clearly suggesting that $1/\xi\lambda \propto T_c^{3/2}$ i.e. that $z=2$. Moreover,  they both decrease roughly linearly with $x$ on either side of the superconducting dome in good agreement with Eq.2. 
 
 Moreover, as pointed out in section II.A, the jump of the specific heat at $T_c$ : $\Delta C_p/T_c \propto [1/\xi\lambda T_c]^2$ and we have also reported on Fig.6a the $T_c\sqrt{\Delta C_p/T_c}$ values (large open circles) deduced from our specific heat measurements. As shown, this quantity consistently follows the expected behavior.  Moreover, one can similarly plot $1/\xi\lambda$, $T_c^{1.5}$ and $T_c\sqrt{\Delta C_p/T_c}$ as a function of doping for Ba(Fe$_x$Co$_{1-x}$)$_2$As$_2$ samples (squares) taking $\lambda$ values from \cite{Williams} and \cite{Luan}, $H'_{c2}$ data from \cite{Vinod} and $T_c\sqrt{\Delta C_p/T_c}$ values from \cite{Budko}. As shown, very similar scalings are obtained in both systems (the $x$ values in Co-doped samples have here been rescaled by a factor $\sim 0.6$ to take into account the difference in the electronic valency between Co and Ni atoms) emphasizing the strong similarity between the two systems. 

This work has been supported by the French National Research Agency, Grant No. ANR-09-Blanc-0211 SupraTetrafer. This work has been supported by the joined PHC grant No.23073WF from the French government and by the Slovak Research and Development Agency, Grant No. SKFR-0024-09.


\begin{references}
\bibitem{anisotropy} preliminary specific heat and Hall probe measurements for $H\|ab$ suggested that $\epsilon\sim 2$ in our samples.
\end{references}
\end{document}